\documentclass[conference,10pt]{IEEEtran}
\IEEEoverridecommandlockouts
\usepackage{amsmath,amssymb,amsfonts}
\usepackage{algorithmic}
\usepackage{graphicx}
\usepackage{textcomp}
\usepackage{xcolor}
\def\BibTeX{{\rm B\kern-.05em{\sc i\kern-.025em b}\kern-.08em
    T\kern-.1667em\lower.7ex\hbox{E}\kern-.125emX}}
\usepackage{cite}
\usepackage{algorithm}
\usepackage{amsthm}
\usepackage{subfigure}
\usepackage{multirow}
\usepackage{stackengine}
\usepackage[colorlinks=true, allcolors=black]{hyperref}

\usepackage{geometry}
\usepackage{array}
\usepackage{mathtools}
\usepackage{cases}
\usepackage{color}
\usepackage{array}
\usepackage{mathtools}
\usepackage{cases}
\begin{document}
\title{Robotic Aerial 6G Small Cells with Grasping End Effectors for mmWave Relay Backhauling }
\author{\parbox{6 in}{\centering Jongyul Lee and Vasilis Friderikos \\
        Centre for Telecommunications Research, Department of Engineering\\
        King's College London, London WC2R 2LS, U.K.\\
        E-mail:\{jongyul.lee, vasilis.friderikos\}@kcl.ac.uk}
}

\maketitle
\begin{abstract}
Deployment of small cells in dense urban areas dedicated to the heterogeneous network (HetNet) and associated relay nodes for improving backhauling is expected to be an important structural element in the design of beyond 5G (B5G) and 6G wireless access networks. A key operational aspect in HetNets is how to optimally implement the wireless backhaul links to efficiently support the traffic demand. In this work, we utilize the recently proposed Robotic Aerial Small Cells (RASCs) that are able to grasp at different tall urban landforms as wireless relay nodes for backhauling. This can be considered as an alternative to fixed small cells (FSCs) which lack flexibility since once installed their position cannot be altered. More specifically, on-demand deployment of RASCs is considered for constructing a millimeter-wave (mmWave) backhaul network to optimize available network capacity using a network flow-based mixed integer linear programming (MILP) formulation. Numerical investigations reveal that for the same required achievable throughput, the number of RASCs required are 25\% to 65\% less than the number of required FSCs. This result can have significant implications in reducing required wireless network equipment (capex) to provide a given network capacity and allows for an efficient and flexible network densification.
\end{abstract}

\IEEEpeerreviewmaketitle
\section{Introduction}
\IEEEPARstart{T}o support the ever increasing data traffic growth of ground users (GUs) in beyond 5G (B5G) or 6G networks, network densification is envisioned to play an important role in augmenting  overall network capacity in the coming years. 
To this end, the deployment of cost-effective wireless backhauling connecting ultra dense small-cell base stations (BSs) will be of fundamental importance in providing the on-demand requirement of the GUs \cite{gao2015mmwave}. 
More specifically, small cell-BSs 
operated at low transmission power can assist conventional macro-BSs to provide extra capacity, connectivity and reliable service, especially when using millimeter-wave (mmWave) spectrum, creating a heterogeneous network (HetNet) to cope with a high data traffic demand increase in urban ultra dense areas \cite{polignano2014inter}. 
To this end, the efficient utilization and operation of mmWave communication has the potential to provide the required gigabit-per-second traffic rate support for wireless backhaul networks in HetNet environments \cite{gao2015mmwave}. 
\begin{figure}[!tb]
\centering
\includegraphics[width=.48\textwidth]{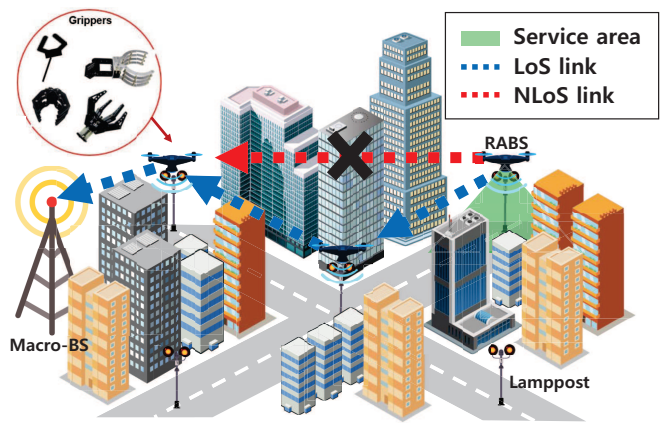}
\caption{\label{fig:overview} An overview of RASC (UAV)-aided B5G mmWave integrated access backhaul network system.}
\end{figure}

Moreover, Unmanned Aerial Vehicles (UAVs) have been considered as a natural extension to the HetNet architecture represented by the small cell-BSs, which offer a highly cost efficient solution to mitigate the expected data increase due to their flexible deployment \cite{frontier}. 
We note that it has been proven technically feasible for the UAV to mount wireless equipment operating as a mmWave aerial BS \cite{semkin2021lightweight}. 
However, since the UAV has inherently limited on-board battery capacity for flying/hovering, sustainable operation as an aerial BS limits the UAV deployment \cite{galkin2019uavs}. Furthermore, the use of a UAV strongly depends on weather conditions such as rain and wind, as well as may cause significant levels of noise pollution in the surrounding area \cite{friderikos2021}.   
To overcome endurance issues of the UAV, a set of previous works have proposed a multiplicity of solutions such as  harvesting solar energy \cite{tuan2021mpc}, applying an free space optics (FSO)-based charging system \cite{wu2020fso}, or using a wired connection as a tether where energy and data feed the UAV whilst hovering \cite{kishk2020aerial}. 
Meanwhile, in \cite{friderikos2021}, \cite{rabs2021}, the concept of Robotic Aerial Small Cell (RASC), which is also called Robotic Aerial Base Station (RABS), is proposed to cope with these issues. It is envisaged that the RASC can provide significant prolonged time of service by grasping in an energy neutral manner at tall urban furniture's such as  lampposts using dexterous grippers. Using such robotic end effectors for grasping, the RASC eliminates the enormous energy consumption required for enabling hovering/flying  of the UAV platform resulting in quickly depleting the limited capacity of the on-board battery \cite{galkin2019uavs}. 
An overview of the RASC-aided mmWave backhaul network above mentioned is illustrated in Fig. \ref{fig:overview}. The figure shows how the flexibility of the RASCs allows them to be deployed where needed including efficient locations of the RASCs that act as relay nodes.

The prior works in \cite{mowla2018energy, huang2019optimal, mcmenamy2019hop, niu2015boosting, yan2021feasibility} proposed the deployment of fixed small cells (FSCs) to create a mmWave backhaul network. 
An energy efficient backhauling optimization problem is proposed to support users and different applications under passive optical network and mmWave backhauling in small cell networks in \cite{mowla2018energy}. 
The authors in \cite{huang2019optimal} create a mmWave backhauling network by selecting cluster heads and optimizing the number of antennas of the macro cell that are dedicated for the backhaul wireless links. 
Also, in \cite{niu2015boosting}, a multi-path multi-hop scheduling scheme is proposed to increase the potential of spatial reuse (concurrent transmission) for mmWave wireless networks. 
The work in \cite{mcmenamy2019hop} develops a mmWave backhaul network framework to maximize network flow for a given topology. Furthermore,  in \cite{yan2021feasibility} the authors investigate the potential of multiple path with relay nodes to maximize throughput for mmWave backhaul networks in urban environments. 
However, to the best of our knowledge, no other works \cite{mowla2018energy, huang2019optimal, mcmenamy2019hop, niu2015boosting, yan2021feasibility} have considered a novel framework explicitly differentiated from \cite{tuan2021mpc, wu2020fso, kishk2020aerial}, where a network flow design is optimized by deploying multiple RASCs serving as small cell-BSs operated as a mmWave backhaul in a dense urban area. One of the major challenges in the deployment of the HetNets is not only to efficiently deploy small cells but also to optimize the location of relay nodes to create high capacity  backhauling  in a given dense urban area \cite{hwang2013holistic}. However, no previous work considered robotic small cells that can be deployed on demand and grasp at different tall urban landforms to be utilized as relay nodes to enhance the wireless backhaul network. 

The key contributions and  novelties of this work can be summarized as follows: 
\begin{itemize}
\item We formulate a network flow optimization problem operated with multiple RASCs deployed from a macro-BS using an integer linear programming (ILP) by considering the mmWave multi-hop backhaul network. The goal of the proposed framework is to minimize the network flow links and RASCs in an efficient manner. 
\item We also augment the formulation by considering the energy consumption of RASCs by using a mixed ILP (MILP), to reduce the total energy consumption when deploying  RASCs.
\item Numerical investigations reveal that RASCs can drastically reduce the amount of network densification since only a quarter to over half of the required number of small cells is employed compared to fixed relay nodes (small cells) to support the same amount of aggregate traffic backhauling rate in the network.
\end{itemize}
\section{System Model}
\subsection{Scenario}
The underlying scenario of the RASC (UAV)-assisted B5G/6G mmWave backhaul network is briefly described below, 
\begin{itemize}
\item A single B5G/6G terrestrial macro-BS is considered in a given  geographical  area and a set of RASCs (multiple RASCs) reside at the macro-BS acting as a depot. The macro-BS is aware regarding areas of increased traffic demand (i.e., large number GUs) that require improvement for the network performance. In that sense, there are clusters of users where RASCs with grasping capabilities can be deployed. 
\item Once the above mentioned clusters are formed (i.e., areas where traffic demand is higher than a given threshold), then RASCs serving as  small cell-BSs for the clusters and  mmWave relay nodes  are dispatched from the depot (i.e., macro-BS) to create a multi-hop wireless network by grasping at suitable lampposts. 
\item When  a RASC arrives at a lamppost, it will utilize its grippers (i.e., robotic arms) to grasp and hence will eliminate any energy consumption for hovering and/or flying. Then, the RASC is on standby to serve and construct the wireless mmWave backhaul links towards the macro-BS as shown in Fig. \ref{fig:overview}. We assume that sufficient radio-frequency (RF) chains are available at the RASC building a link so that the link is used by multiple flows \cite{xiao2017millimeter, mcmenamy2019hop}. 
Since the urban environment scenario with many obstacles for the blockage effect such as high-rise buildings is assumed, Line-of-Sight (LoS) connection (link) is only desirable for mmWave communication. Also, the service last for a given duration $t$. 
\end{itemize}
\subsection{Network Model}
Without loss of generality, we assume that all RASCs moved between nodes fly with a constant velocity $v$ and stay for the duration $t$ at similar lampposts in shape, hence all RASCs are at the same height when they are serving users and/or used as relay nodes. 
With that in mind, the aforementioned scenario in the UAV-assisted B5G/6G backhaul network (also called the UAV-assisted B5G/6G relay network) is modeled as an undirected graph $\mathcal{G}$ = $(\mathcal{V},\mathcal{E})$, where  $\mathcal{V}$=$\{1,2,\cdots,N_{V}\}$ denotes a set of nodes
\footnote{Hereafter, the term 'node' and 'lamppost' are used interchangeably. Also, the term 'node' relates to a RASC acting as serving small cell-BS or a RASC acting as a relay for backhauling.}
(i.e., both lampposts and a macro-BS (depot)), the elements of which are included in a 2D Cartesian coordinate system, i.e., the horizontal plane.
$\mathcal{E}$=$\{(i ,j):i,j \in \mathcal{V}, i \neq j\}$ is a set of links. $\mathcal{V}=\mathcal{V}_{e} \cup \mathcal{V}_{b}$, where $\mathcal{V}_{e}=\{1,2,\cdots,N_{E}\}$ denotes a set of hotspot areas where GUs require service and $\mathcal{V}_{b}=\{1,2,\cdots,N_{B}\}$ represents a set of nodes (lampposts and the macro-BS) where the RASC stays for the service to connect and create the backhaul links. 
Also, $\mathcal{F}$=$\{\gamma_{\text{th}}^{1},\gamma_{\text{th}}^{2},\cdots,\gamma_{\text{th}}^{N_{E}}\}$ denotes a set of flows that is the required throughput (service) for GUs and we define $\gamma_{\text{th}}^{f}[t]$ as the flow $f$ at the duration $t$.
Thus, $|\mathcal{V}_{e}|$=$|\mathcal{F}|$, where $|.|$ is defined as the cardinality of the set. 
Last but not least, $\mathcal{K}$=$\{1,2,\cdots,N_{K}\}$ denotes a set of RASCs that perform a role as both a serving RASC and a relay RASC.
\subsection{Transmission Model}
We consider the utilization of mmWave communications at 28 GHz frequency. 
The mmWave links are realized by employing fixed beamforming with high-gain antenna arrays mounted on the RASC where each antenna array covers a 90$^\circ$ sector in the horizontal plane.
The mmWave signals are generally less prone to mutual interference due to the directional nature of the propagation beamforming resulted from the use of narrow beamwidth antennas \cite{yan2021feasibility, mohamed2020relay}. 
Furthermore, the strong directional signals (with the blockage effect and the many obstacles such as high-rise buildings in the urban environment) reduce the overall impact of interference.
For our scenario; the urban wireless backhaul network, nodes (at the top of the lampposts that the RASC grasps) are located on a height from 5m to 12m (which is the typical range for the size of lampposts in UK), which is unlikely for a physical object to disturb their links, i.e., Non-LoS (NLoS) links.
Hence, we assume that no more than two physical links along the constructed path interfere with each other for simplicity \cite{polignano2014inter}. That is, we only consider Signal to Noise Ratio (SNR) and LoS condition for the link.
For a detailed mmWave link capacity generation as a function of the distance between nodes, we apply the macroscopic pathloss as discussed in \cite[Table I]{akdeniz2014millimeter} in order to calculate the beamforming gains. 

\subsection{Energy Consumption Model for Rotary Wings of a UAV}
We adopt a RASC controlled with rotary wings, the constant velocity $v$, and grippers (robotic arms). Hence, a widely-used energy consumption model is applied and plugged into the proposed framework. The energy consumption is divided into three parts as follows.
\begin{enumerate}
\item \textit{Propulsion Energy:} The propulsion power with a function of the velocity $v$ in meter per second (m/s) is approximately given by $P_{\text{trav}}(v)$ described in \cite[Fig. 12]{zeng2019energy}. Subsequently, the propulsion energy in Joule (J) to fly between nodes is denoted as $E_{\text{trav}}^{ij}=P_{\text{trav}}(v)t_{\text{trav}}^{ij}, \forall i,j (i \neq j) \in \mathcal{V}$, where $t_{\text{trav}}^{ij}$ represents the traveling time that the RASC spends flying, i.e., $t_{\text{trav}}^{ij}=d_{ij}/v$, where $d_{ij}$ is the Euclidean distance between nodes. 
\item \textit{Grasping Energy:} The grasping power of RASC depends on its size and weight, as well as the type of the gripper \cite{friderikos2021}. The electromagnetic solenoid-based grippers can be deemed as suitable for attaching to ferromagnetic surfaces such as lampposts \cite{nedungadi2019design}. Hence, the grasping energy consumed during the duration $t$ is given by $E_{\text{grasp}} = {P}_{\text{grasp}}[t]$, where ${P}_{\text{grasp}}$ is the grasping power. We note that this can be considered as an upper bound while the RASC is perching since energy neutral grippers of the RASC that could be applied require no energy consumption for hovering and/or flying. 
\item \textit{Communication Energy:} The communication energy of the RASC is given by $E_{\text{comm}}=(P_{0}+\eta_{p}P_{\text{TX}})[t]$, where $P_{0}$ is the minimum active power, $\eta_{p}$ is a linear transmission factor and $P_{\text{TX}}$ is the transmission power as discussed in \cite{hoyhtya2018review}.
\end{enumerate}
\section{Problem Formulation}
Based on the discussed scenario and aforementioned system model, we provide a proposed ILP formulation for network flow path and location optimization of robotic aerial platforms with grasping end effectors acting as small cell-BSs or relay nodes. To begin with, a binary decision variable for the network flow of the proposed model can be defined as follows,
\begin{equation}
\label{var_y}
y_{ijf}=
\begin{cases}
1,  & \text{if flow $f$ uses a link}  \\
&\text{between a node $i$ and a node $j$,} \\
0,  & \text{otherwise.}
\end{cases}
\end{equation}
We then derive several constraints to create the network flow path in the wireless backhaul network as follows,
\begin{subequations}
\begin{equation}
\begin{aligned} 
\label{flow1}
&\sum_{j \in \mathcal{V}} y_{ijf} = 1,&&\forall i \in \mathcal{V}_{e},&&\forall f \in \mathcal{F},\\ 
\end{aligned}
\end{equation}
\begin{equation}
\begin{aligned} 
\label{flow2}
&\sum_{i (i\neq 0)\in \mathcal{V}} y_{i0f} = 1,&&\forall f\in \mathcal{F}, \\ 
\end{aligned}
\end{equation}
\begin{equation}
\begin{aligned} 
\label{flow3}
&\sum_{j (i\neq j)\in \mathcal{V}_{b}} y_{ijf} = \sum_{j (i\neq j)\in \mathcal{V}_{b}} y_{jif},&&\forall i \in \mathcal{V}_{b}\backslash \{0\},&& \forall f\in \mathcal{F}, \\
\end{aligned}
\end{equation}
\end{subequations}
where the index 0 denotes the macro-BS (depot). The constraints in (\ref{flow1}) and (\ref{flow2}) represent that the flow $\gamma_{\text{th}}^{f}$ starts to use the link from the very beginning area (i.e., the hotspot area) and ends to the macro-BS. The constraint in (\ref{flow3}) guarantees flow conservation of the network flow. 

Next, we also define a binary decision variable for multiple RASCs to connect the flow and provide relevant constraints as follows. 
\begin{equation}
\label{var_x}
x_{ik}=
\begin{cases}
1,  & \text{if a RASC $k$ stays at a node $i$ }  \\
&\text{for the network flow,} \\
0,  & \text{otherwise,}
\end{cases}
\end{equation}
\begin{subequations}
\begin{equation}
\begin{aligned} 
\label{rasc_loc1}
&\sum_{k\in \mathcal{K}}x_{ik} \leq 1,&&\forall i \in \mathcal{V}\backslash\{0\},\\
\end{aligned}
\end{equation}
\begin{equation}
\begin{aligned} 
\label{rasc_loc2}
&\sum_{i\backslash\{0\} \in \mathcal{V}}x_{ik} \leq 1,&&\forall k \in \mathcal{K}, \\
\end{aligned}
\end{equation}
\begin{equation}
\begin{aligned} 
\label{rasc_loc3}
&\sum_{k\in \mathcal{K}}x_{ik} \geq y_{ijf},&&\forall i \in \mathcal{V}\backslash\{0\},&&j \in \mathcal{V}_{b},&&\forall f \in \mathcal{F}.\\
\end{aligned}
\end{equation}
\end{subequations}
The constraints in (\ref{rasc_loc1}) and (\ref{rasc_loc2}) ensure that only one RASC $k$ stays at the node $i$ when it creates and connects the link for the flow. The constraint in (\ref{rasc_loc3}) allows multi-flow to traverse through the link ($i,j$) with a RASC $k$ as a node $i$. 

For a given deployment of the network flow and corresponding optimum locations of RASCs, we also model constraints on link capacity as given by, 
\begin{subequations}
\begin{equation}
\begin{aligned} 
\label{cap1}
&S_{ij}^{\text{eff}} =\text{min} [\log_{2}(1+10^{0.1*(\text{SNR}-\alpha)}),S_{ij}^{\text{max,eff}}],~\forall i,j \in \mathcal{V},\\
\end{aligned}
\end{equation}
\begin{equation}
\begin{aligned} 
\label{cap2}
&c_{ij}=B \cdot S_{ij}^{\text{eff}},&&\forall i,j\in \mathcal{V},\\
\end{aligned}
\end{equation}
\begin{equation}
\begin{aligned} 
\label{cap3}
&\sum_{f \in \mathcal{F}} \gamma_{\text{th}}^{f}[t] y_{ijf} \leq c_{ij},&&\forall i, j\in \mathcal{V},\\
\end{aligned}
\end{equation}
\end{subequations}
where $\alpha$ is a loss factor in dB obtained empirically, and $B$ is the bandwidth of the channel. The constraint in (\ref{cap1}) denotes spectral efficiency on each link by the Shannon capacity in bit/s/Hz discussed in \cite{akdeniz2014millimeter}. Observe that for a practical network, $S^{\text{max,eff}}$ protects infinite values which causes the unrealistic scenario. The constraints in (\ref{cap2}) and (\ref{cap3}) impose that the sum of all flows is restricted under the capacity $c$. 

Based on the above decision variables in (\ref{var_y}) and (\ref{var_x}), and specific constraints, the optimization model for the flow of the backhaul network with multiple RASCs can be formulated as follows,
\begin{equation}
\begin{aligned} 
\label{obj1}
\text{(P1): }~\min_{\substack{\mathbf{X},\mathbf{Y}}} \sum_{i \in \mathcal{V}}\sum_{j\in \mathcal{V}}\sum_{f\in \mathcal{F}} y_{ijf} +\sum_{i \in \mathcal{V}}\sum_{k \in \mathcal{K}} x_{ik} \\
\end{aligned}
\end{equation}
\begin{equation}
\begin{aligned}
\label{p1}
\text{s. t.} \:\:\: ~(\ref{flow1})-(\ref{flow3}),~(\ref{rasc_loc1})-(\ref{rasc_loc3}),~(\ref{cap1})-(\ref{cap3}),
\end{aligned}
\end{equation}
\begin{equation}
\begin{aligned} 
\label{var_x1}
x_{ik} \in \{0,1\},&&\forall i\in \mathcal{V}\backslash\{0\},&&\forall k \in \mathcal{K},
\end{aligned} 
\end{equation}
\begin{equation}
\begin{aligned} 
\label{var_y1}
y_{ijf} \in \{0,1\},&&\forall i,j (i\neq j)\in \mathcal{V},&&\forall f \in \mathcal{F},
\end{aligned} 
\end{equation}
where $\mathbf{X} \triangleq \{x_{ik} \big| \, \forall i \in \mathcal{V},~\forall k \in \mathcal{K} \}$, $\mathbf{Y} \triangleq \{y_{ijf} \big| \, \forall i,j \in \mathcal{V},~\forall f \in \mathcal{F} \}$. 
The objective function in (\ref{obj1}) is to minimize the number of hops for the network flow while minimizing the number of RASCs. 

Furthermore, in order to prevent the use of inefficient energy consumption of RASCs when they are dispatched from the macro-BS, we derive an additional continuous variable and constraints on the energy consumption of the RASC as follows,
\begin{equation}
\begin{aligned} 
\label{var_e}
E_{ik}^{\text{total}} \geq 0, \ & \text{energy consumption of a RASC $k$} \\
& \text{when it is activated at a node $i$,} \\
\end{aligned} 
\end{equation}
\begin{subequations}
\begin{equation}
\begin{aligned} 
\label{e1}
&E_{jik}^{\text{fly}}=x_{ik}E_{\text{trav}}^{ji},&&\forall i,j \in \mathcal{V},&&\forall k \in \mathcal{K},\\ 
\end{aligned}
\end{equation}
\begin{equation}
\begin{aligned} 
\label{e2}
&E_{ik}^{\text{act}}=x_{ik}E_{\text{grasp}}+\sum_{f \in \mathcal{F}}y_{ijf}E_{\text{comm}},&&\forall i, j\in \mathcal{V},&&\forall k \in \mathcal{K},\\ 
\end{aligned}
\end{equation}
\begin{equation}
\begin{aligned} 
\label{e3}
&E_{ik}^{\text{total}}=E_{jik}^{\text{fly}}+E_{ik}^{\text{act}},&&\forall i,j \in \mathcal{V},&&\forall k \in \mathcal{K},\\ 
\end{aligned}
\end{equation}
\end{subequations}
where $E_{jik}^{\text{fly}}$ is the energy consumption when the RASC travels from the node $j$ to the node $i$, $E_{ik}^{\text{act}}$ is the energy consumption for flowing into all flows when the RASC stays at a node $i$ for the duration $t$, and $E_{ik}^{\text{total}}$ is the sum of all the energy consumption of a RASC. Worth pointing out that in the constraint (\ref{e2}), the $E_{\text{comm}}$ linearly scales depending on the number of flows, i.e., RF chains \cite{hoyhtya2018review, mcmenamy2019hop}.

With the continuous variable in (\ref{var_e}) and the constraints (\ref{e1})$-$(\ref{e3}), we propose a MILP formulation for network flow of the proposed model as follows, 
\begin{equation}
\begin{aligned} 
\label{obj2}
\text{(P2): }~\min_{\substack{\mathbf{E}, \mathbf{X}, \mathbf{Y}}} &\sum_{i \in \mathcal{V}}\sum_{j\in \mathcal{V}}\sum_{f\in \mathcal{F}} y_{ijf}+\sum_{i\in \mathcal{V}}\sum_{k\in \mathcal{K}} x_{ik} \\&+\sum_{i\in \mathcal{V}}\sum_{k\in \mathcal{K}}E_{ik}^{\text{total}}\\
\end{aligned}
\end{equation}
\begin{equation}
\begin{aligned}
\label{p2}
\text{s. t.} \:\:\: &~(\ref{flow1})-(\ref{flow3}),~(\ref{rasc_loc1})-(\ref{rasc_loc3}),~(\ref{cap1})-(\ref{cap3}),~(\ref{var_x1}),(\ref{var_y1}),\\&(\ref{e1})-(\ref{e3}),
\end{aligned}
\end{equation}
\begin{equation}
\begin{aligned} 
\label{var_e1}
E_{ik}^{\text{total}} \geq 0,&&\forall i\in \mathcal{V}\backslash\{0\},&&\forall k \in \mathcal{K},
\end{aligned} 
\end{equation}
where $\mathbf{E} \triangleq \{E_{ik}^{\text{total}} \big| \, \forall i \in \mathcal{V},~\forall k \in \mathcal{K} \}$.
The objective function in (\ref{obj2}) is to minimize the number of hops for the network flow while minimizing the number of RASCs and the energy consumption of RASCs. We note that the problem (P1) and (P2) are the combinatorial optimization problem as a multi-commodity low-cost flow problem, which is the well-known NP-complete problem \cite{ahuja1993network}.

\begin{table}[!tb]
\caption{\label{tab:parameters} Simulation Parameters.} 
\centering
\begin{tabular}{l|c}
\hline
\textbf{Parameter}&\textbf{Value} \\
\hline
\hline
Deployment area & Manhattan grid model \\ 
& (150m$\times$150m)\\
\hline
Number of blocks (buildings) & 9 (30m$\times$30m)\\
\hline
Number of lampposts (nodes)& 16\\
including a macro-BS  ($N_{V}$) & \\
\hline
Number of required flows above & 1, 2, 3 \\
the threshold at respective hotspot area ($N_{E}$)& \\
\hline
Maximum number of RASCs ($N_{K}$) & 15 \\
\hline
Required throughput in the hotspot area ($\gamma_{\text{th}}$)& [0.5, 3] bps/Hz\\
\hline
Duration ($t$)& 1800 s \\ 
\hline
Carrier frequency & 28 GHz \\ 
\hline
Bandwidth ($B$) & 1 GHz \\ 
\hline
Minimum active power ($P_{0}$)& 6.8 W \cite{hoyhtya2018review}\\
\hline
Linear transmission factor ($\eta_{p}$)& 4 W \cite{hoyhtya2018review}\\ 
\hline
Transmit power ($P_{\text{TX}}$)& 24 dBm \cite{liu2018performance} \\ 
\hline
Grasping power ($P_{\text{grasp}}$)& 10 W \cite{nedungadi2019design}\\ 
\hline
Propulsion power ($P_{\text{trav}}$)& \cite{zeng2019energy} in detail \\ 
\hline
Velocity ($v$)& 10.21 m/s \cite{zeng2019energy}\\ 
\hline
Maximum spectral efficiency ($S^{\text{max,eff}}$) & 4.8 bps/Hz \cite{mcmenamy2019hop}\\ 
\hline
Loss factor ($\alpha$) & 3 dB \cite{mcmenamy2019hop}\\ 
\hline
\end{tabular}
\end{table}

\begin{figure}[!tb]
\centering
\includegraphics[width=.47\textwidth]{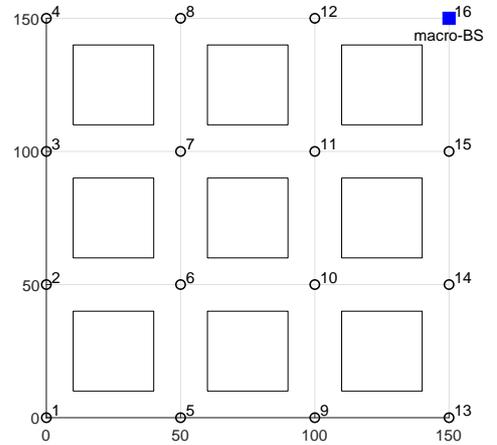}
\caption{\label{fig:map} RASC deployment in the Manhattan grid model.}
\vspace{-0.0cm}
\end{figure}
\vspace{0.5cm}
\section{Numerical Investigations}
In this section, a wide set of Monte Carlo based numerical investigations are presented to shed light on the effectiveness and potential achievable gains  of the proposed RASCs acting as relay nodes for creating mmWave backhauling compared to FSCs. Simulations are based on MATLAB using \textquoteleft Intlinprog\textquoteright~solver. 

Table \ref{tab:parameters} summarizes the parameters used for performance evaluation and Fig. \ref{fig:map} illustrates an overview of the deployment. 
For a typical urban scenario, simulations are conducted in a Manhattan grid model aiming at a city (i.e., a typical downtown urban environment), the size of which is 150m$\times$150m. 
The size of the street blocks (buildings) that restrict propagation  are 30m$\times$30m and the street width is 10m (the edge part) or 20m (the street along the intersection). 
Also, we consider 15 lampposts, at 10m height which are uniformly distributed at a distance of 50m apart along the intersection of the streets. These lampposts constitute the set of candidate locations for hosting the RASCs. In other words, these lampposts are the locations where the RASCs will fly to and grasp via the embedded dexterous end effectors eliminating in that sense energy consumption for hovering and/or flying. 
A macro-BS is located on the corner of the very top right as shown in Fig. \ref{fig:map}. 
We vary the number of hotspots in the given area creating traffic flows denoted as $N_{E}$. Those hotspots are randomly distributed to the vicinity of lampposts in the given area.
The maximum deploying number of RASCs ($N_{K}$) is 15, which is the same number depending on the number of lampposts ($N_{V}$) so that they are able to be assigned, if needed, at all available locations. 
We assume, without loss of generality, that all required achieved throughput at hotspot areas is identical, i.e., $\gamma_{\text{th}}^{f}=\gamma_{\text{th}}$.
We utilize the velocity $v$ where the RASC consumes the minimum
energy with the maximum endurance as detailed in \cite{zeng2019energy} and the transmit power $P_{\text{TX}}$ in \cite{liu2018performance} as reference points  in all simulations presented hereafter.

Due to space limitations, we do not discuss the energy consumption of RASCs for (P1) and (P2) solutions. 

\begin{figure}[!tb]
\centering
\includegraphics[width=.50\textwidth]{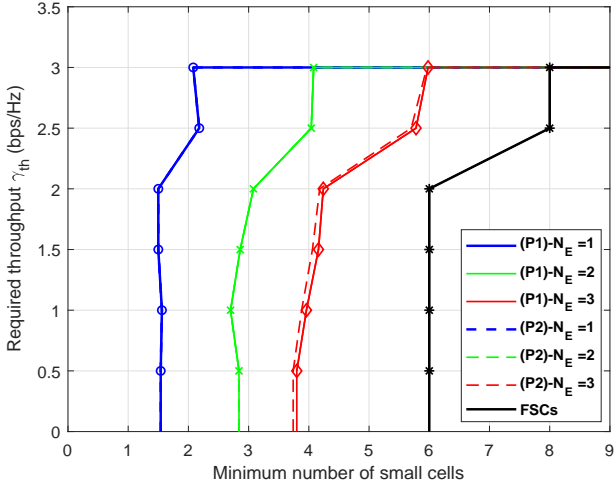}
\caption{\label{fig:required} The minimum number of small cells of the proposed (P1), (P2) solutions, and FSCs with varying $N_{E}$ and $\gamma_{\text{th}}$. }
\end{figure}
\begin{figure}[!tb]
\centering
\begin{subfigure}[(left figure) (P1) solution for the typical case. (right figure) FSCs.]{
\includegraphics[width=.49\textwidth]{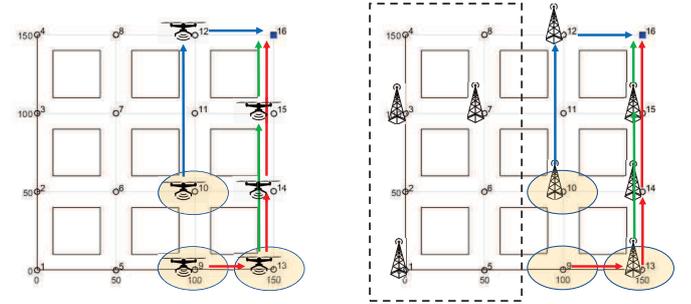}
\label{fig:toy1}}

\end{subfigure}

\begin{subfigure}[(left figure) (P1) solution for a worst case scenario. (right figure) FSCs.]{
\includegraphics[width=.49\textwidth]{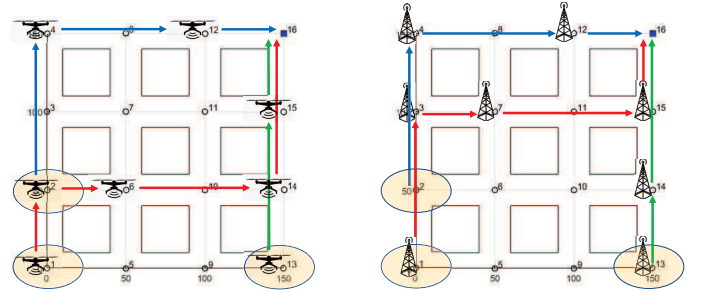}
\label{fig:toy2}}

\end{subfigure}
\caption{\label{fig:toy} Toy examples of (P1) solution and FSCs for the case of $N_{E}=3$ and $\gamma_{th}=3$ bps/Hz.}
\end{figure}
Fig. \ref{fig:required} shows the minimum number of small cells with varying $\gamma_{\text{th}}$ and $N_{E}$. We compare the RASC deployment of the proposed (P1) and (P2) solutions in the case of $N_{E}=1$ to 3 with the FSCs deployment. The FSCs as infrastructure with wireless equipment may cause a highly expensive installation cost \cite{jafari2015small, bouras2014financing}.
Moreover, they are in essence deployed as immovable, but set to be connected to cover the given whole area for enabling the traffic flow from the hotspot areas to the macro-BS regardless of $N_{E}$ for the wireless backhaul. 
Worth to note that all schemes steeply increase the number of small cells at $\gamma_{\text{th}}=2.5$ bps/Hz due to the link capacity. 
It is observed that our proposed optimal scheme (P1) and (P2) use more efficient implementation in constructing backhaul links than the FSCs throughout the entire simulations. On average the number of RASCs required are  2.08, 4.08 and 5.98 for the case of $N_{E}$ range from 1 to 3 respectively when end-users at hotspots require a minimum rate of 3 bps/Hz  ($\gamma_{\text{th}}$).
However, for the same achievable rate the number of fixed cells ranges from 6 to 8 and this is regardless of increase of the value $N_{E}$ (i.e., the number of hotspots). This means that the gains range from 25\% to 65\% in terms of reducing the actual small cells required for the same capacity. This is due to the fact that their location is fixed in the given area,
and their position is selected based on a minimum rate to be achieved. However, when hotspots are created in locations which are not in the vicinity of the FSC location and require the highest rate, then the required number of FSCs is significantly increased. 
Note that the (P2) solution gains the same value as the (P1) solution when $N_{E}=1$ and 2. Meanwhile, the (P2) solution with $N_{E}=3$ gains slightly lower minimum number of RASCs than the (P1) solution. This is because the (P2) solution eliminates (minimizes) inefficient links, i.e., results in reduction of the number of RASCs, under the constraints in (\ref{e1})$-$(\ref{e3}) in terms of the energy consumption of RASCs deployment.

Fig. \ref{fig:toy} shows toy examples of the (P1) solution and the FSCs when $N_{E}=3$ and $\gamma_{th}=3$ bps/Hz. 
All left figures are the result of the (P1) solution whilst all right figures are the result of the FSCs. 
In the figures, the yellow circle indicates hotspot areas and the red, green and blue arrows indicate  flow direction. 
In Fig. \ref{fig:toy1}, only 6 RASCs are deployed for the flow starting from the three hotspot areas for the typical case. 
Whereas, 5 FSCs are deployed for the same three hotspot areas whilst additional 3 FSCs cover the rest of the given urban area as described in the black dotted box, i.e., the total number of the FSCs is 8.
In the left of Fig. \ref{fig:toy2}, 8 RASCs are deployed for a worst-case scenario in the location of hotspots. Whereas, 8 FSCs with different allocations are also deployed to cover the given entire urban area. Hence, we note that the maximum number of RASCs required for the deployment does not exceed the number of FSCs in the given same urban area.

\vspace{0.5cm}
\section{Conclusions and Future Work}
This paper has presented a novel optimization framework to compute an optimal network flow implemented by multiple RASCs serving users at hotspots areas and acting as relay nodes to create a wireless mmWave backhaul network. 
The proposed framework allows for the creation of the backhaul links so that the required flow is optimized from the hotspot areas to the macro-BS when there are multiple RASCs perched at lampposts. 
The proposed framework outperforms the deployment of the conventional FSCs for supporting the same throughput demand by providing 25\% to 65\% lower number of RASCs than that of the FSCs. This implies that due to their efficient on-demand deployment and the fact that they do not consume energy for hovering and/or flying while perching, RASCs can be seen as a considerably attractive alternative to FSC installation. 

As future work, we aim to devise novel algorithms to create a joint optimization of RASCs with dexterous end effectors and integrated access and backhaul mmWave spectrum allocation. A meticulous analysis of energy consumption for deploying RASCs with neutral grasping capabilities will also form part of further research. 
\vspace{0.5cm}
\bibliographystyle{IEEEtran}
\bibliography{reference}

\begin{thebibliography}{10}
\providecommand{\url}[1]{#1}
\csname url@samestyle\endcsname
\providecommand{\newblock}{\relax}
\providecommand{\bibinfo}[2]{#2}
\providecommand{\BIBentrySTDinterwordspacing}{\spaceskip=0pt\relax}
\providecommand{\BIBentryALTinterwordstretchfactor}{4}
\providecommand{\BIBentryALTinterwordspacing}{\spaceskip=\fontdimen2\font plus
\BIBentryALTinterwordstretchfactor\fontdimen3\font minus
  \fontdimen4\font\relax}
\providecommand{\BIBforeignlanguage}[2]{{%
\expandafter\ifx\csname l@#1\endcsname\relax
\typeout{** WARNING: IEEEtran.bst: No hyphenation pattern has been}%
\typeout{** loaded for the language `#1'. Using the pattern for}%
\typeout{** the default language instead.}%
\else
\language=\csname l@#1\endcsname
\fi
#2}}
\providecommand{\BIBdecl}{\relax}
\BIBdecl

\bibitem{gao2015mmwave}
Z.~Gao \emph{et~al.}, ``Mmwave massive-mimo-based wireless backhaul for the 5g
  ultra-dense network,'' \emph{IEEE Wireless communications}, vol.~22, no.~5,
  pp. 13--21, 2015.

\bibitem{polignano2014inter}
M.~Polignano \emph{et~al.}, ``The inter-cell interference dilemma in dense
  outdoor small cell deployment,'' in \emph{2014 IEEE 79th vehicular technology
  conference (VTC Spring)}.\hskip 1em plus 0.5em minus 0.4em\relax IEEE, 2014,
  pp. 1--5.

\bibitem{frontier}
I.~Bor-Yaliniz and H.~Yanikomeroglu, ``The new frontier in ran heterogeneity:
  Multi-tier drone-cells,'' \emph{IEEE Communications Magazine}, vol.~54,
  no.~11, pp. 48--55, 2016.

\bibitem{semkin2021lightweight}
V.~Semkin \emph{et~al.}, ``Lightweight uav-based measurement system for
  air-to-ground channels at 28 ghz,'' \emph{arXiv preprint arXiv:2103.17149},
  2021.

\bibitem{galkin2019uavs}
B.~Galkin, J.~Kibilda, and L.~A. DaSilva, ``Uavs as mobile infrastructure:
  Addressing battery lifetime,'' \emph{IEEE Communications Magazine}, vol.~57,
  no.~6, pp. 132--137, 2019.

\bibitem{friderikos2021}
V.~Friderikos, ``{Airborne urban microcells with grasping end fffectors: A game
  changer for 6G networks?}'' \emph{arXiv preprint arXiv:2105.09230}, 2021.

\bibitem{tuan2021mpc}
H.~D. Tuan \emph{et~al.}, ``Mpc-based uav navigation for simultaneous
  solar-energy harvesting and two-way communications,'' \emph{IEEE Journal on
  Selected Areas in Communications}, vol.~39, no.~11, pp. 3459--3474, 2021.

\bibitem{wu2020fso}
D.~Wu, X.~Sun, and N.~Ansari, ``An fso-based drone charging system for
  emergency communications,'' \emph{IEEE Transactions on Vehicular Technology},
  vol.~69, no.~12, pp. 16\,155--16\,162, 2020.

\bibitem{kishk2020aerial}
M.~Kishk, A.~Bader, and M.-S. Alouini, ``Aerial base station deployment in 6g
  cellular networks using tethered drones: The mobility and endurance
  tradeoff,'' \emph{IEEE Vehicular Technology Magazine}, vol.~15, no.~4, pp.
  103--111, 2020.

\bibitem{rabs2021}
Y.~Liao and V.~Friderikos, ``{Optimal Deployment and Operation of Robotic
  Aerial 6G Small Cells with Grasping End Effectors},'' \emph{arXiv preprint
  arXiv:2110.03794}, 2021.

\bibitem{mowla2018energy}
M.~M. Mowla \emph{et~al.}, ``Energy efficient backhauling for 5g small cell
  networks,'' \emph{IEEE Transactions on Sustainable Computing}, vol.~4, no.~3,
  pp. 279--292, 2018.

\bibitem{huang2019optimal}
P.-H. Huang and K.~Psounis, ``Optimal backhauling for dense small-cell
  deployments using mmwave links,'' \emph{Computer Communications}, vol. 138,
  pp. 32--44, 2019.

\bibitem{mcmenamy2019hop}
J.~McMenamy \emph{et~al.}, ``Hop-constrained mmwave backhaul: Maximising the
  network flow,'' \emph{IEEE Wireless Communications Letters}, vol.~9, no.~5,
  pp. 596--600, 2019.

\bibitem{niu2015boosting}
Y.~Niu, C.~Gao, Y.~Li, D.~Jin, L.~Su, and D.~Wu, ``Boosting spatial reuse via
  multiple-path multihop scheduling for directional mmwave wpans,'' \emph{IEEE
  Transactions on Vehicular Technology}, vol.~65, no.~8, pp. 6614--6627, 2015.

\bibitem{yan2021feasibility}
Y.~Yan, Q.~Hu, and D.~M. Blough, ``Feasibility of multipath construction in
  mmwave backhaul,'' in \emph{2021 IEEE 22nd International Symposium on a World
  of Wireless, Mobile and Multimedia Networks (WoWMoM)}.\hskip 1em plus 0.5em
  minus 0.4em\relax IEEE, 2021, pp. 81--90.

\bibitem{hwang2013holistic}
I.~Hwang, B.~Song, and S.~S. Soliman, ``A holistic view on hyper-dense
  heterogeneous and small cell networks,'' \emph{IEEE Communications Magazine},
  vol.~51, no.~6, pp. 20--27, 2013.

\bibitem{xiao2017millimeter}
M.~Xiao \emph{et~al.}, ``Millimeter wave communications for future mobile
  networks,'' \emph{IEEE Journal on Selected Areas in Communications}, vol.~35,
  no.~9, pp. 1909--1935, 2017.

\bibitem{mohamed2020relay}
E.~M. Mohamed \emph{et~al.}, ``Relay probing for millimeter wave multi-hop d2d
  networks,'' \emph{IEEE Access}, vol.~8, pp. 30\,560--30\,574, 2020.

\bibitem{akdeniz2014millimeter}
M.~R. Akdeniz \emph{et~al.}, ``Millimeter wave channel modeling and cellular
  capacity evaluation,'' \emph{IEEE journal on selected areas in
  communications}, vol.~32, no.~6, pp. 1164--1179, 2014.

\bibitem{zeng2019energy}
Y.~Zeng, J.~Xu, and R.~Zhang, ``Energy minimization for wireless communication
  with rotary-wing uav,'' \emph{IEEE Transactions on Wireless Communications},
  vol.~18, no.~4, pp. 2329--2345, 2019.

\bibitem{nedungadi2019design}
A.~S. Nedungadi and M.~Saska, ``Design of an active-reliable grasping mechanism
  for autonomous unmanned aerial vehicles,'' in \emph{International Conference
  on Modelling and Simulation for Autonomous Systems}.\hskip 1em plus 0.5em
  minus 0.4em\relax Springer, 2019, pp. 162--179.

\bibitem{hoyhtya2018review}
M.~H{\"o}yhty{\"a}, O.~Apilo, and M.~Lasanen, ``Review of latest advances in
  3gpp standardization: D2d communication in 5g systems and its energy
  consumption models,'' \emph{Future Internet}, vol.~10, no.~1, p.~3, 2018.

\bibitem{ahuja1993network}
R.~K. Ahuja, T.~L. Magnanti, and J.~B. Orlin, \emph{Network flows: Theory,
  algorithms, and applications}.\hskip 1em plus 0.5em minus 0.4em\relax
  Pearson, 1993.

\bibitem{liu2018performance}
C.~Liu \emph{et~al.}, ``Performance analysis for practical unmanned aerial
  vehicle networks with los/nlos transmissions,'' in \emph{2018 IEEE
  International Conference on Communications Workshops (ICC Workshops)}.\hskip
  1em plus 0.5em minus 0.4em\relax IEEE, 2018, pp. 1--6.

\bibitem{jafari2015small}
A.~H. Jafari \emph{et~al.}, ``Small cell backhaul: challenges and prospective
  solutions,'' \emph{EURASIP Journal on Wireless Communications and
  Networking}, vol. 2015, no.~1, pp. 1--18, 2015.

\bibitem{bouras2014financing}
C.~Bouras, V.~Kokkinos, and A.~Papazois, ``Financing and pricing small cells in
  next-generation mobile networks,'' in \emph{International Conference on
  Wired/Wireless Internet Communications}.\hskip 1em plus 0.5em minus
  0.4em\relax Springer, 2014, pp. 41--54.

\end{thebibliography}

\end{document}